\documentclass[12pt]{iopart}

\usepackage[figuresright]{rotating}
\begin{document}


\title[]{On the Galactic Center Being the Main Source of Galactic Cosmic
Rays as Evidenced by Recent Cosmic Ray and Gamma Ray Observations}

\author{Yi-Qing Guo, Zhao-Yang Feng, Qiang Yuan, Cheng Liu, Hong-Bo Hu}

\address{Key Laboratory of Particle Astrophysics, Institute of High
Energy Physics, Chinese Academy of Science, Beijing 100049, P.R.China}
\ead{fengzy@ihep.ac.cn}

\begin{abstract}
{We revisit the idea that the Galactic center (GC) is the dominant
source of Galactic cosmic rays (GCRs), based on a series of new
observational evidence. A unified model is proposed to explain the
new phenomena of GCRs and $\gamma$-rays simultaneously. The GCRs
are thought to be accelerated during past activities of the GC.
The pair production process of GCRs in the strong radiation field
due to the GC activity is responsible for the knee structure of the
cosmic ray spectra. A fraction of $e^+e^-$ produced by pair production
interactions, can be reaccelerated in the induced bipolar jets and
be transported into the halo, leaving the Fermi $\gamma$-ray bubbles
and WMAP microwave haze as the remnant signal. Finally, the CRs diffuse
in the bulge could further interact with the interstellar medium to
produce low energy $e^+e^-$. After cooling down, these positrons may
annihilate to produce the 511 keV line emission as discovered by INTEGRAL.
}
\end{abstract}

\maketitle

\section{Introduction}

~~~~~~The origin of cosmic rays (CRs) has been a mystery since their discovery
in 1912. CRs have a nearly featureless power-law spectrum with a spectral
index of about $-3$ from energies of $\sim 10^{9}$ to $\sim 10^{20}$ eV.
However, detailed measurements revealed several subtle structures in the
CR spectrum at $\sim4\times10^{15}$ eV (knee), $\sim4\times10^{17}$ eV
($2$nd knee), $\sim 5\times10^{18}$ eV (ankle), and $\sim 6\times10^{19}$
eV (Greisen-Zatsepin-Kuzmin cut-off) \cite{Horandel07}.
The study of these structures is of great importance for understanding
the origin, propagation and interaction of CRs.

It is generally believed that supernova remnants (SNRs) are the sources
of Galactic CRs (GCRs), based on the simple argument that the power by
supernovae is sufficient to sustain the total power of GCRs
\cite{1934Baade,1964Ginzburg}. But there are still some open questions
for SNRs as the dominant sources of GCRs, such as the isotopic
abundance, radial gradient of CRs deduced from diffuse $\gamma$-rays,
and the maximum attainable energy of particles \cite{2005JPhG,2009Natur.460}.
The evidence from $\gamma$-ray observations is also indirect and
inconclusive. Therefore we could have enough motivation to discuss
the origin of GCRs beyond SNRs.

Instead of discussing the stellar level sources in the Galaxy, the
Galactic center (GC) as powered by the accretion of the supermassive
black hole, could be a natural candidate of CR origin. Although the
GC is relatively quiet nowadays, there is large amount of evidence
suggesting that the GC may have been active $\sim 10^7$ years ago 
\cite{1971A&A....13..405V,1974ApJ...188..489S,2007JPhG...34.1813E}.
Historically, here have been many discussions on the possibility
that the GC is the dominant source of GCRs as a result of its activity
\cite{1981AZh....58..959P,1981ICRC....2..344S,1983JPhG....9.1139G}.

There are great progresses in the measurements of GCRs and $\gamma$-rays
in recent years. First, owing to the improved energy resolution, a very
sharp knee structure was found by many air shower array experiments
\cite{2007APh....28...58C,2007NuPhS.165...74K,2008JPhG...35k5201G,
2008ApJ...678.1165A,2009APh....31...86A,2009NJPh...11f5008I}. This result
seems to favor the single source model as proposed in
\cite{1997JPhG...23..979E,2009arXiv0906.3949E}. Second, several high
precision measurements led to the discovery of both the positron fraction
excess \cite{2009Natur.458..607A} and the total $e^+e^-$ spectral excess
\cite{2008Natur.456..362C,2009PhRvL.102r1101A} from $O(10)$ GeV to TeV.
It was also shown that the $e^+e^-$ spectra have a cutoff at several
TeV \cite{2008PhRvL.101z1104A,2009A&A...508..561A}. The result seems
also to favor the single source model proposed in \cite{2009ApJ...700L.170H}.
Since the observational anisotropy of CRs is very weak ($10^{-4}\sim10^{-3}$),
the single source should not be very close to us. The GC is one of the
potential candidates \cite{2009arXiv0911.3034H}. Furthermore, the
discoveries of the 511 keV line emission and $\gamma$-ray bubbles in the
GC region indicate that the GC may indeed have past activities.
Based on these new observational data, it is time to revisit the idea that
the GC is the main source of GCRs.

In this work we describe a unified model to explain the recent observations
of CRs and $\gamma$-rays, based on the past activity of the GC.

\section{Model}
~~~~~~During the active phase of the GC, the accretion of stars and
gases by the supermassive black hole is very efficient to provide
high enough power for particle acceleration, jet launch and propagation.
The GC activities can produce shocks and accelerate CRs to high energies
\cite{Cheng:2006ch,2001MNRAS.321..433B} in scales from $\sim10^{-3}$
pc \cite{2008ApJ...685L..23A} to kpc \cite{2012ApJ...753...61S}.
The heating of the accretion disk makes it filled with thermal photons.
Jets could also be produced due to the accretion events.

\begin{figure}[!htb]
\centering
\includegraphics[width=0.8\textwidth,angle=-0]{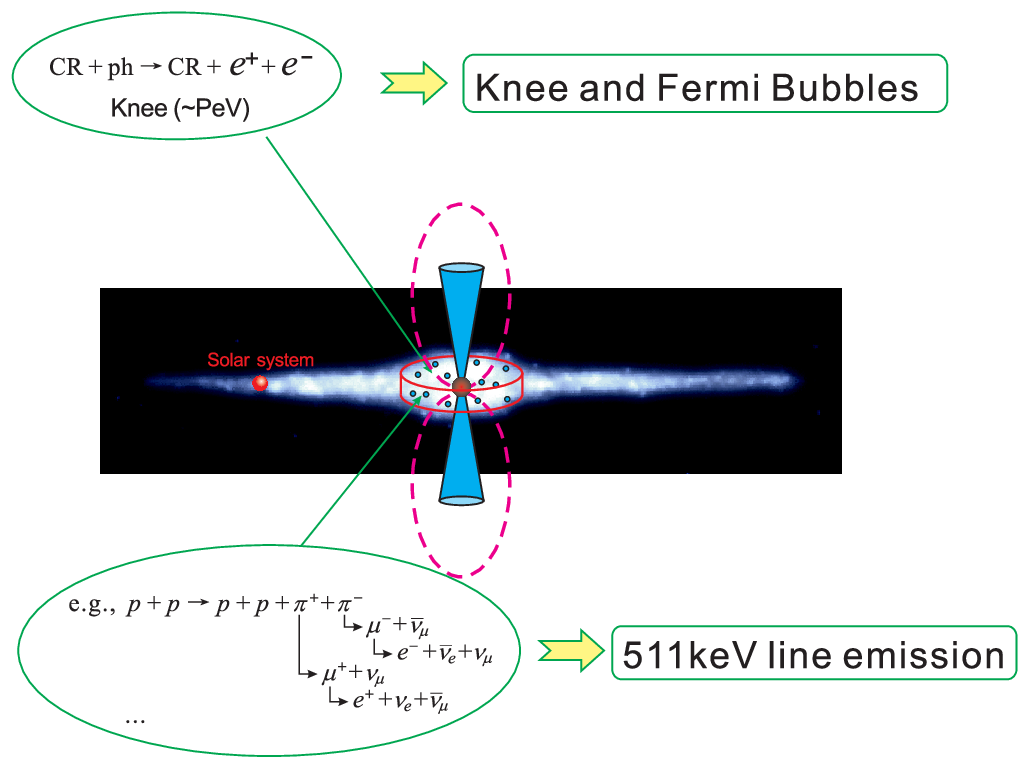}
\caption{{\bf The cartoon to describe the unified model. The GCRs are
accelerated in the most central region around the black hole, during
past activities. Pair production processes of GCRs in the strong radiation
field occur and are responsible for the knee structure of the spectra.
A fraction of the produced $e^+e^-$ enters into the halo and radiate
to form the bubbles. The inelastic interactions between the CRs and
the gas in the bulge and disk further produce the 511 keV line emission.
}}
\label{GC_Picture}
\end{figure}

In the above mentioned environment, the following physical processes are
expected to occur. First we may have the primary acceleration of CRs
around the black hole. The accelerated nuclei can then escape from
the source region and enter the disk filled with background field.
The interaction between CR nuclei and the background photons can lead
to energy loss of CRs, which could result in the formation of the knee
\cite{2009ApJ...700L.170H}. The interaction will produce $e^+e^-$ pairs,
which can be partially
transported into the halo through the jets. Those electrons and positrons
can then radiate to form the multi-wavelength haze/bubble. Finally the
CRs diffuse in the bulge could further interact with the interstellar
medium to produce low energy $e^+e^-$. After cooling down, these positrons
may annihilate to produce the 511 keV line emission. The cartoon to
describe the basic picture of the model is shown in Fig.\ref{GC_Picture}.
In the following we will discuss more details about the three aspects,
the origin of the knee, the multi-wavelength haze/bubble and the 511
keV emission respectively.

\subsection{Origin of the Knee}

Ref. \cite{2009ApJ...700L.170H} proposed a model to simultaneously
explain the knee and the $e^+e^-$ excess, incorporating pair production
interactions between CRs and the ambient photon field. It was further shown
that the irregular structures of the CR spectra around the knee region and
the Galactic ``B component'' could also be well explained in this scenario
\cite{2010ScChG..53..842W}. As a consequence, one set of parameters in
the mode of \cite{2009ApJ...700L.170H} also favors the single source
model. It indicates that there might be a single source with relatively
stable properties (during the CR acceleration period) that is responsible
for GCRs.

In \cite{2009ApJ...700L.170H}, a supernova-pulsar system was proposed as
a possible candidate for such a source. Though it is not impossible, it
seems non-trivial for such a system to satisfy the conditions needed to
produce the knee of the CRs \cite{2011ASTRA...7..179E}. Alternative
candidate sources may include micro-quasars or the Galactic center (GC).
The latter seems to be an especially attractive option
\cite{2009arXiv0911.3034H}. As proposed by many studies, the capture of
stars or accretion of gas by the central supermassive black hole can
produce shock and accelerate particles \cite{Cheng:2006ch}.

The density of background radiation field at GC region is the key point
whether the knee structure of CR spectra can be formed. As shown in
\cite{2009ApJ...700L.170H}, the photon column density should be
$\sim 10^{30}$ cm$^{-2}$\footnote{ Note for such a photon field
the photodisintegration and pion-production processes will be too
strong. Since these two processes have higher threshold energy, they
could be suppressed if the maximum energy of CRs do not exceed much
beyond the knee.}.  Without firm observational evidence of the
size of such kind of interactions, we take $1$ pc as an illustration.
The infrared-optical photon density inside $1$ pc of the GC was
about $10^4-10^5$ eV/cm$^{3}$ in a flaring disk and dust
\cite{1992ApJ...387..189D,1996A&ARv...7..289M}. As an estimate we adopt
a value of $5\times10^4$ eV/cm$^{3}$. For photons with typical energy
$\sim$eV, the photon number density is about $5\times10^4$ cm$^{-3}$.
The diffusion velocity of CRs in the knee region is estimated to
be the order of $10^{-3}$ of the light velocity, according to the measured
anisotropy \cite{2003ICRC....1..183A}. Therefore, the time for CRs to
diffuse out of such a region is about $10^3$ yr. So the photon column
density which CRs can encounter is $n c \tau \sim 10^{26}$ cm$^{-2}$,
which is 4 orders of magnitude lower than that required. However, it is
possible that the photon luminosity could be much higher during the
active phase than at present. For example, the observation of infrared
radiation from other galaxies showed that when the nucleus was in the
active phase the infrared luminosity could be as high as $10^{44}-10^{47}$
erg/s, which is $2-5$ orders of magnitude higher than the present value
of our Galaxy, $\sim10^{42}$ erg/s \cite{1981AZh....58..959P}.
Therefore it is possible that the background photon density could be 4
orders of magnitude higher during the active phase, and the condition to
form the knee could be satisfied. Under these circumstances, the total
energy of background photon is estimated to be $\sim 10^{53}$ erg,
which is close to the accretion power of one solar mass.

\subsection{Fermi bubbles: possible relics of past GC activity}

~~~~~~If GC indeed plays a significant role to produce the Galactic CRs,
we may expect the existence of some relics of the past activity of GC.
``Fermi bubbles'', the new observational evidence, may be such kind of
relics of the past GC activity.

Thanks to the high performance of Fermi $\gamma$-ray telescope, a large
scale, extended $\gamma$-ray excess in the GC direction was discovered
\cite{2010ApJ...717..825D}, which was then revealed to be two giant
$\gamma$-ray bubbles \cite{2010ApJ...724.1044S}. The Fermi bubbles are
symmetric with respect to the Galactic plane, extending $\sim 50$ degrees
in latitude and $\sim 40$ degrees in longitude. They are spatially
correlated with the WMAP haze observed in the $20-60$ GHz band
\cite{2004ApJ...614..186F,2008ApJ...680.1222D}, and the edges of the
bubbles are also found to be coincident with features in the ROSAT $1.5-2$
keV X-ray maps \cite{1997ApJ...485..125S}. Recently, the PLANCK collaboration
confirmed the microwave haze found in WMAP data \cite{2012arXiv1208.5483P}.
Several models are proposed to explain the Fermi bubbles
\cite{2011arXiv1110.0834G,2011ApJ...731L..17C,2011PhRvL.106j1102C,
2011arXiv1103.0055G,2011arXiv1110.5436I,2011PhRvL.107i1101M,
2011MNRAS.415L..21Z}, most of which are based on the GC activity in the past.

The bubbles are found to have a hard $\gamma$-ray spectrum between 1 and
100 GeV, with a power law index $\sim -2$. The $\gamma$-ray spectrum can
be well reproduced by the inverse Compton scattering (ICS) process of
power-law distributed electrons with index $-2\sim -2.5$
\cite{2010ApJ...724.1044S}, taking into account the cosmic microwave
background, infrared and optical
background radiation. In addition, the calculated synchrotron radiation
can reproduce the radio haze flux, assuming that the magnetic field is of
the order of 10 $\mu$G. However, this electron spectrum has difficulty
to explain the observed low energy drop below 1 GeV. In order to solve
this problem, an electron population with limited energy range was
proposed. Based on these facts, Su et al. \cite{2010ApJ...724.1044S}
concluded that the bubbles are most likely created by a large episode of
energy injection in the GC in the last $10^{7}$ years through an accretion
event in the center of supermassive black hole, a nuclear starburst or
some other energetic event.

The locally measured energy density of CRs is about $1$ eV cm$^{-3}$.
Giving that the volume of the Galactic disk is about $\pi(20\,{\rm kpc})^2
(0.2\,{\rm kpc}) \sim 10^{67}$ cm$^3$, the total energy of CRs is
about $10^{55}$ erg. Assuming the pre-propagated spectrum of GCRs
is $\propto E^{-2.0}$, the total energy of GCRs above the knee
($E\sim$PeV) is approximately $10^{54}$ erg. Such an energy will be
mostly converted into $e^+e^-$ through the pair production interactions.
According to \cite{2010ApJ...724.1044S}, the Fermi bubbles has an age
on the order of $10^{7}$ yr. It is to say that the luminosity of $e^+e^-$
is estimated to be $10^{39}-10^{40}$ erg/s. The total luminosity of the
Fermi bubbles in $1-100$ GeV is estimated to be about $4\times10^{37}$ erg/s
\cite{2010ApJ...724.1044S}, which is much smaller than the above estimated
value.  That is to say, a small fraction of the produced $e^+e^-$ could
be enough to generate the Fermi bubbles\footnote{ It can be seen below
that the luminosity of the synchrotron emission is even higher than the
$\gamma$-ray emission, which means a higher fraction of the produced
$e^+e^-$ is necessary.}

It should be noted that in the strong background radiation field,
$e^+e^-$ may cool down very efficiently. The timescale is estimated
to be the order of $10^{-2}$ year for TeV electrons. However, it
is expected that there should also be acceleration in the jets and/or
in the inner 1 pc region around the GC. The acceleration of the
active galactic nuclei jet could be very efficient, such as minutes
to hours \cite{2007ApJ...664L..71A}, which may possibly compensate
the cooling of the $e^+e^-$.

\begin{figure}[!htb]
\centering
\includegraphics[width=0.45\textwidth,angle=0]{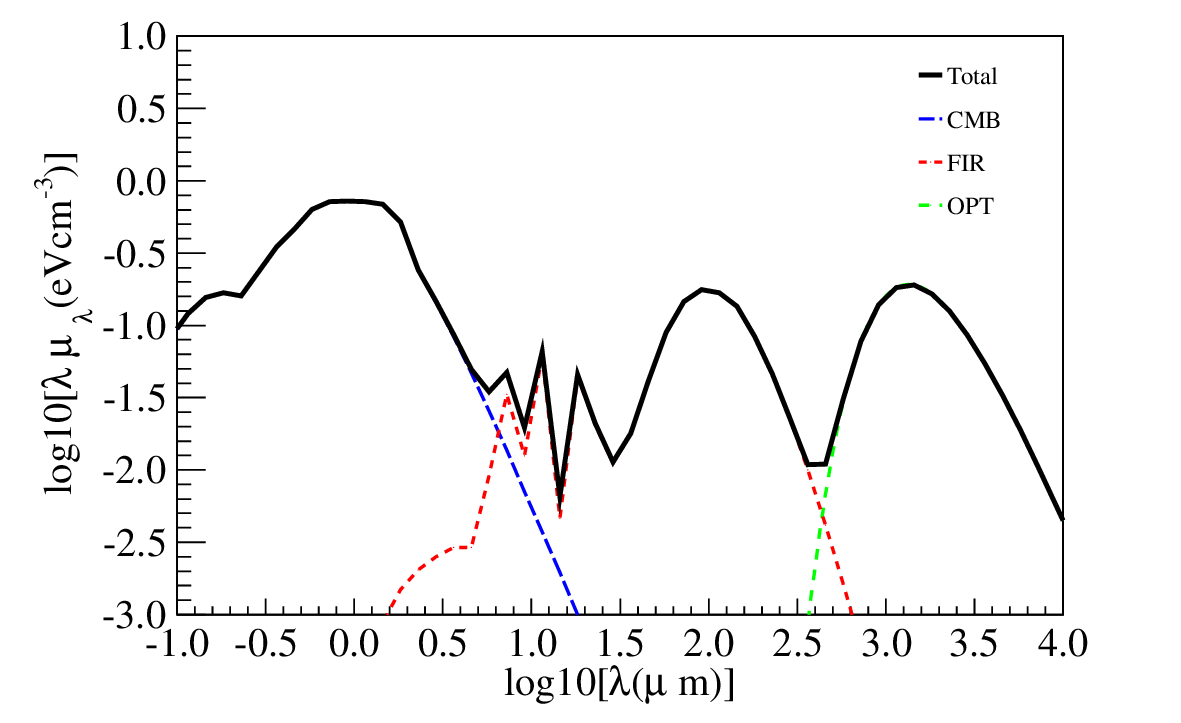}
\includegraphics[width=0.45\textwidth,angle=0]{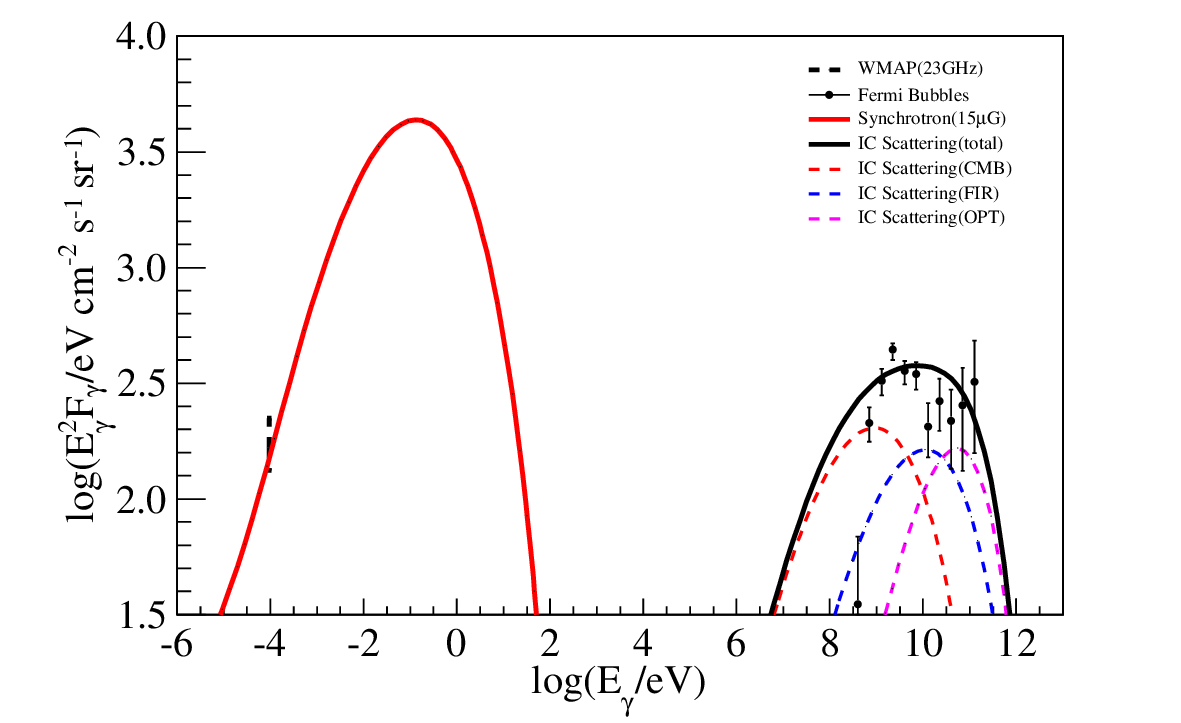}
\caption{Left: the ISRF at $(R,z)=(0\,{\rm kpc},\,4\,{\rm kpc})$, taken from GALPROP package. Right: The calculated spectrum of ICS
$\gamma$-rays and synchrotron radiation originating from a re-accelerated
electron spectrum generated through CR-photon pair production interactions.
The line of sight direction is chosen to be $l=0^{\circ}$ and $b=25^{\circ}$.
The data points representing the Fermi bubbles and WMAP haze are taken from
Table 3 and Fig. 23 of \cite{2010ApJ...724.1044S}.}
\label{fig-FermiBubble-WMAPHaze}
\end{figure}

The $e^+e^-$ spectrum used to calculate the synchrotron and ICS spectra
is taken from \cite{2009ApJ...700L.170H},
which can explain the $e^+e^-$ excesses observed by PAMELA/ATIC/Fermi.
For the interstellar radiation field (ISRF) model we adopt that reported
in Porter \& Strong \cite{2005ICRC....4...77P}, in which a new calculation
based on the modelings of star and dust distributions, the scattering,
absorption and re-emission of the stellar light by dust, was carried out.
The ISRF model showed good agreement with the observational data
\cite{2006ApJ...640L.155M}, and was implemented in the public CR
propagation code GALPROP \cite{1998ApJ...509..212S}. Here we adopt the
ISRF intensity at $R=0$ and $z=4$ kpc. The energy spectrum of the ISRF
is shown in the left panel of Fig. \ref{fig-FermiBubble-WMAPHaze}.
Three major components, optical from stars, far-infrared from dust
and the cosmic microwave background (CMB), are clearly shown.

The right panel of Fig. \ref{fig-FermiBubble-WMAPHaze} shows the resulting
synchrotron and ICS spectra by the $e^+e^-$. The magnetic field is assumed
to be $B=15\mu$G, and the Klein-Nishina cross section of ICS is adopted.
The spectra of WMAP synchrotron haze and Fermi ICS bubbles are consistently
generated.

\subsection{511 keV line emission}

~~~~~~It is natural to expect a possible connection of the GC origin of
CRs with the 511 keV line emission as reported by several experiments
\cite{Johnson:1972,2005A&A...441..513K,Jean:2005af,Weidenspointner:2008zz},
although the most popular model of 511 keV emission is the decay of
radioactive isotope\cite{2011RvMP...83.1001P}.
The 511 keV line emission indicates the existence of non-relativistic
positrons in the GC region. The hadronic interactions of GCRs with the
ambient gas could be one potential source of these positrons
\cite{Cheng:2006ch,Ramaty:1970ra,Totani:2006ff}. We make an order of
magnitude estimate of the power of electrons. The total number of CR
protons is about $\sim10^{58}$, for a local number density of $\sim10^{-9}$
cm $^{-3}$ and the volume of the Galactic disk $\sim 10^{67}$ cm$^3$.
Considering that the size of the Galactic bulge is $\sim 1$ kpc, the
typical path length that a particle travels from the GC to outside of
the bulge should be $\sim 10^3$ kpc, for diffusion coefficient
$D\sim 5\times 10^{28}$ cm$^2$ s$^{-1}$. Assuming that the number
density of ISM nuclei in the bulge is $1$ cm$^{-3}$, and the inelastic
cross section of $p-p$ scattering is several tens of mb, the average
number of collisions for one CR proton before traveling out of the bulge
is $\sim 0.1$. Thus, the total number of positrons is $\sim 10^{57}$.
Assuming the cooling time of positrons is about $10^7$ years, which
corresponds to the ionization and Coulomb losses in an ISM with density
of $1$ cm$^{-3}$ for a $100$ MeV positron \cite{1998ApJ...509..212S},
the cooled positron production rate is $3\times10^{42}$ s$^{-1}$, which
is comparable with the rate $10^{43}$ s$^{-1}$ as implied from the flux
of 511 keV $\gamma$-ray line \cite{2005A&A...441..513K}.

However, it was pointed out that the diffuse $\gamma$-ray constrained
positron production rate would be not more than a few percent of the
positron rate suggested by the 511 keV emission data
\cite{Cheng:2006ch,2005A&A...441..513K,2008ApJ...682..400P}. This
problem can be solved in a non-stationary scenario that the GC was
in active phases in the past and the positron production rate would
be much higher than that determined by the current diffuse $\gamma$-ray
flux \cite{Cheng:2006ch,2008ApJ...682..400P}.

\section{Conclusion and discussion}

~~~~~~In this work we proposes that the GC is the major source of GCRs.
There is evidence to show the past activity of the GC. Particle
acceleration can take place during the violent phase of the GC.
Also it is expected the existence of strong radiation field around
the GC. Thus an efficient $e^+e^-$ pair production interactions
between GCRs and the ambient photons might be responsible for the
the knee of the CR spectra \cite{2009ApJ...700L.170H}. A fraction of
$e^+e^-$ produced by  pair production interactions, can be reaccelerated
in the jets and escape into the halo. The ICS and synchrotron radiation
of these $e^+e^-$ may possibly explain the observed Fermi bubbles and WMAP haze.

Even though the jet can transport the $e^+e^-$ very efficiently into the halo,
the propagation of these particles from the jet to the whole bubble is
still an open question. If we adopt the diffusion velocity of $10^{-3}c$ as in the disk, the electrons/positrons need
$10^7$ years to travel $\sim5$ kpc. It is much larger than the typical cooling time
($<10^{6}$ years \cite{2010ApJ...724.1044S}) of TeV electrons. However, a faster transportation of the
particles in the halo is possible.

Because of the lack of knowledge about the Galaxy magnetic
field, phenomenological model is generally used to study the propagation of
GCRs in the Galaxy. As we know the halo is much larger than the disk.
But the average density of the medium GCRs travel is $\sim0.3$ cm$^{-3}$
while the disk density is about $1$ cm$^{-3}$ \cite{HaloPropagation}.
So the trapping time of GCRs in the halo is only about two times longer
than in in the disk. We can infer that the propagation velocity in the
halo is much faster than in the disk. It is possible to conclude that the
stochastic magnetic field is much smaller in the halo than in the disk.
We need to investigate the particle transportation in the regular halo
magnetic field \cite{2001SSRv...99..243B} to study propagation of particles
in the halo, instead of the uniform diffusion in the Galaxy. Therefore
the $e^+e^-$ may fill in the whole bubble within the cooling time through
the fast propagation. However, the detailed model goes beyond the scope of
the current work.

The anisotropy of GCRs may still an open question in the GC scenario
of the origin of GCRs. Still the different propagation parterns in the
disk and the halo may lead to different results of the expected anisotropy.
We leave the discussion of it in future works.

Finally we should note that in the jet the background electrons should also
be accelerated together with the $e^+e^-$ produced through CR-photon
interactions. In such a case the total electron spectrum used to calculate
the bubble/haze emission might be different from what adopted in this work.
Without loss of generality, we may expect the background electron spectrum
to be a power-law spectrum $\sim E^{-2}$ with a cutoff, which does not
differ much from that we use here. The basic results in this work should
not change significantly.

\section*{Acknowledgements}

We thank Shaoxia Chen for helpful discussion, Yigang Xie, Amanda Maxham,
Ann Meng Zhou and Hanguo Wang for useful comments on the manuscript.
This work is supported by the Ministry of Science and Technology of
China, Natural Sciences Foundation of China (Nos. 10725524,10773011 and
11135010), and the Chinese Academy of Sciences (Nos. KJCX2-YW-N13,
KJCX3-SYW-N2, GJHZ1004).


\section*{References}
\begin{thebibliography}{10}
\bibitem{Horandel07}
  H{\"o}randel, J. R. 2007, Modern Physics Letters A, 22, 1533
\bibitem{1934Baade}
Baade, W. \& Zwicky, F. 1934, Phys. Rev., 46, 76
\bibitem{1964Ginzburg}
Ginzburg, V. L. \& Syrovatskii, S. I. 1964, The Origin of Cosmic Rays (New York:Macmillan)
\bibitem{2005JPhG}
Hillas, A.~M. 2005, Journal of Physics G Nuclear Physics, 31, 95
\bibitem{2009Natur.460}
Butt, Y. 2009, Nature, 460, 701
\bibitem{1971A&A....13..405V}
van der Kruit, P.~C. 1971, A\&A, 13, 405
\bibitem{1974ApJ...188..489S}
Sanders, R.~H., \& Prendergast, K.~H. 1974, ApJ, 188, 489
\bibitem{2007JPhG...34.1813E}
Erlykin, A.~D., \& Wolfendale, A.~W. 2007, Journal of Physics G Nuclear Physics, 34, 1813
\bibitem{1981AZh....58..959P}
Ptuskin, V.~S., \& Khazan, Y.~M. 1981, AZh, 58, 959
\bibitem{1981ICRC....2..344S}
Said, S.~S., Wolfendale, A.~W., Giler, M., \& Wdowczyk, J. 1981 Proc. 17\MakeLowercase{$^{th}$} Int. Cosmic Ray Conf. (Paris) Vol 2 p 344
\bibitem{1983JPhG....9.1139G}
Giler, M. 1983, Journal of Physics G Nuclear Physics, 9, 1139
\bibitem{2007APh....28...58C}
Chilingarian, A., et al. 2007, Astroparticle Physics, 28, 58
\bibitem{2007NuPhS.165...74K}
Korosteleva, E.~E., Prosin, V.~V., Kuzmichev, L.~A., \& Navarra, G. 2007, Nuclear Physics B Proceedings Supplements, 165, 74
\bibitem{2008JPhG...35k5201G}
Garyaka, A.~P., et al. 2008, Journal of Physics G Nuclear Physics, 35, 115201
\bibitem{2008ApJ...678.1165A}
Amenomori, M., et al. 2008, ApJ, 678, 1165
\bibitem{2009APh....31...86A}
Apel, W.~D., et al. 2009, Astroparticle Physics, 31, 86
\bibitem{2009NJPh...11f5008I}
Ivanov, A.~A., Knurenko, S.~P., \& Sleptsov, I.~Y. 2009, New Journal of Physics, 11, 065008
\bibitem{1997JPhG...23..979E}
Erlykin, A.~D., \& Wolfendale, A.~W. 1997, Journal of Physics G Nuclear Physics, 23, 979
\bibitem{2009arXiv0906.3949E}
Erlykin, A.~D., \& Wolfendale, A.~W. 2009, arXiv:0906.3949
\bibitem{2009Natur.458..607A}
Adriani, O., et al. 2009, Nature, 458, 607
\bibitem{2008Natur.456..362C}
Chang, J., et al. 2008, Nature, 456, 362
\bibitem{2009PhRvL.102r1101A}
Abdo, A.~A., et al. 2009, Physical Review Letters, 102, 181101
\bibitem{2008PhRvL.101z1104A}
Aharonian, F., et al. 2008, Physical Review Letters, 101, 261104
\bibitem{2009A&A...508..561A}
Aharonian, F., et al. 2009, A\&A, 508, 561
\bibitem{2009ApJ...700L.170H}
Hu, H.-B., et al. 2009, ApJ, 700, L170
\bibitem{2009arXiv0911.3034H}
Hu, H, Proc. 31\MakeLowercase{$^{st}$} Int. Cosmic Ray Conf. ({\L}\'{o}d\'{z} 2009) pp 87-94(arXiv:0911.3034)
\bibitem{Cheng:2006ch}
Cheng, K. S., et al. 2006, ApJ, 645, 1138
\bibitem{2001MNRAS.321..433B}
Bell, A.R., Lucek, S.G., 2001, MNRAS, 314, 65
\bibitem{2008ApJ...685L..23A}
Albert, J., Aliu, E.,Anderhub, H., et al.\ 2008, ApJL, 685, L23
\bibitem{2012ApJ...753...61S}
Su, M., et al., 2012, ApJ, 753,61
\bibitem{2010ScChG..53..842W}
Wang, B., et al. 2010, Science in China G: Physics and Astronomy, 53, 842
\bibitem{2011ASTRA...7..179E}
Erlykin, A.~D., Wibig, T. \& Wolfendale, A.~W. 2011, Astrophysics and Space Sciences Transactions, 7, 179
\bibitem{1992ApJ...387..189D}
Davidson, J. A., et al. 1992, ApJ, 387, 189
\bibitem{1996A&ARv...7..289M}
Mezger, P. G., Duschl, W. J. $\&$ Zylka, R., 1996, A$\&$AR, 7, 289
\bibitem{2003ICRC....1..183A}
Aglietta, M., et al., 2003, ICRC, 183
\bibitem{2010ApJ...717..825D}
Dobler, G., Finkbeiner, D.~P., Cholis, I., Slatyer, T., \& Weiner, N. 2010, ApJ, 717, 825
\bibitem{2010ApJ...724.1044S}
Su, M., Slatyer, T.~R., \& Finkbeiner, D.~P.\ 2010, ApJ, 724, 1044
\bibitem{2004ApJ...614..186F}
Finkbeiner, D.~P. 2004, ApJ, 614, 186
\bibitem{2008ApJ...680.1222D}
Dobler, G., \& Finkbeiner, D.~P. 2008, ApJ, 680, 1222
\bibitem{1997ApJ...485..125S}
Snowden, S.~L., et al. 1997, ApJ, 485, 125
\bibitem{2012arXiv1208.5483P}
Ade, P.A.R., 2012, arXiv:1208.5483v1
\bibitem{2011arXiv1110.0834G}
Guo, F. et al. 2012, ApJ, 756, 182
\bibitem{2011ApJ...731L..17C}
Cheng, K. S., et al. 2011, ApJ, 731, L17
\bibitem{2011PhRvL.106j1102C}
Crocker, R. M., \& Aharonian, F. 2011, Physical Review Letters, 106, 101102
\bibitem{2011arXiv1103.0055G}
Guo, F., \& Mathews, W. G. 2011, arXiv:1103.0055
\bibitem{2011arXiv1110.5436I}
Istomin, Y. N. 2011, arXiv:1110.5436
\bibitem{2011PhRvL.107i1101M}
Mertsch, P., \& Sarkar, S. 2011, Physical Review Letters, 107, 091101
\bibitem{2011MNRAS.415L..21Z}
Zubovas, K., King, A.~R. \& Nayakshin, S. 2011, MNRAS, 415, 21
\bibitem{2007ApJ...664L..71A}
Aharonian, F., Akhperjanian, A.~G., Bazer-Bachi, A.~R., et al. 2007, ApJL, 664, L71
\bibitem{2005ICRC....4...77P}
Porter, T. A., \& Strong, A. W. 2005, ICRC, 4, 77
\bibitem{2006ApJ...640L.155M}
Moskalenko, I. V., Porter, T. A., \& Strong, A. W. 2006, ApJ, 640, L155
\bibitem{1998ApJ...509..212S}
Strong, A.~W., \& Moskalenko, I.~V. 1998, ApJ, 509, 212
\bibitem{Johnson:1972}
Johnson, W. N., et al. 1972, ApJ, 172, L1
\bibitem{2005A&A...441..513K}
Kn{\"o}dlseder, J., et al. 2005, A\&A, 441, 513
\bibitem{Jean:2005af}
Jean, P., et al.  2005, A\& A, 445, 579
\bibitem{Weidenspointner:2008zz}
Weidenspointner, G., et al. 2008, Nature, 451, 159
\bibitem{2011RvMP...83.1001P}
Prantzos, N. et al, 2011, RvMP, 83, 1001
\bibitem{Ramaty:1970ra}
Ramaty, R. et al. 1970, Journal of Geophysical Research, 75, 1141
\bibitem{Totani:2006ff}
Totani, T. 2006, PASJ, 58, 965
\bibitem{2008ApJ...682..400P}
Porter, T. et al. 2008, ApJ, 682, 400
\bibitem{HaloPropagation}
Maurin, D., et al., 2001, ApJ, 555, 585
\bibitem{2001SSRv...99..243B}
Beck, R. 2001, Space Science Reviews, 99, 243
\end{thebibliography}
\end{document}